\documentclass[aip,jap,onecolumn,showpacs,superscriptaddress]{revtex4-1}  
\usepackage{graphicx}  
\usepackage{float}
\usepackage{dcolumn}   
\usepackage{bm}        
\usepackage{amssymb}   
\usepackage{amsmath}
\usepackage{isomath}
\usepackage[usenames, dvipsnames]{color}
\usepackage{xcolor}
\usepackage{soul}
\usepackage{physics}
\usepackage[normalem]{ulem}

\setcitestyle{super}

\begin{document}

\title{Elastic Wannier-Stark Ladders and Bloch Oscillations in 1D Granular Crystals}

\author{Xiaotian Shi}
\affiliation{Aeronautics and Astronautics, University of Washington, Seattle, WA 98195, USA}
\author{Rajesh Chaunsali}
\affiliation{Aeronautics and Astronautics, University of Washington, Seattle, WA 98195, USA}
\author{Ying Wu}
\affiliation{Aeronautics and Astronautics, University of Washington, Seattle, WA 98195, USA}
\affiliation{Department of Astronautic Science and Mechanics, Harbin Institute of Technology, Harbin, Heilongjiang 150001, China}
\author{Jinkyu Yang}
\thanks{Corresponding author: jkyang@aa.washington.edu}
\affiliation{Aeronautics and Astronautics, University of Washington, Seattle, WA 98195, USA}


\begin{abstract}
We report the numerical and experimental study of elastic Wannier-Stark Ladders and Bloch Oscillations in a  tunable one-dimensional granular chain consisting of cylindrical particles. The Wannier-Stark Ladders are obtained by tuning the contact angles to introduce a gradient in the contact stiffness along the granular chain. These ladders manifest as resonant modes localized in the space. When excited at the corresponding resonant frequencies, we demonstrate the existence of time-resolved Bloch Oscillations. The direct velocity measurements using Laser Doppler Vibrometry agree well with the numerical simulation results. We also show the possibility of further tailoring these Bloch Oscillations by numerical simulations. Such tunable systems could be useful for applications involving the spatial localization of elastic energy. 
\end{abstract}


\maketitle

\section{INTRODUCTION}
The study of energy localization in materials has increasingly drawn attention of researchers in recent years.\citep{L1,L2} On these lines, granular crystals, which are tightly-packed arrangements of granular particles interacting as per nonlinear contact law, have been attractive due to their tunability.\citep{g1,g2,g3} Such systems are fertile test beds of studying elastic wave propagation as a result of intermixing effects of dispersion, nonlinearity, defect, and disorder. The localization of wave---in the form of the spatially localized linear or nonlinear resonant modes---has thus been shown in the studies related to breathers,\citep{g4, g5, g6} defects,\citep{g7, g8} topological interfaces,\citep{g9} and disorder.\cite{g10, g11, g12, g13}  


Here, we explore a new way of achieving wave localization in granular systems by the formation of the elastic Wannier-Stark Ladders (WSLs) and Bloch Oscillations (BOs). BOs were originated from quantum mechanics referring to the spatially localized oscillations of electrons in a periodic potential with an external applied electrical field. After those were predicted by Bloch and Zener,\citep{1} this phenomenon remained controversial for more than 60 years.\citep{2} A.G. Chynoweth \textit{et al}. observed WSLs in crystals with an applied electric field and verified the Bloch-Zener model.\citep{3} These WSLs referred to the equally-spaced electron energy bands and represented the frequency-domain counterpart of BOs. After the initial experimental verifications of electric BOs in semiconductor superlattice,\citep{4} analogous systems have been studied and demonstrated for plasmonic waves,\citep{5} light waves,\citep{2,6,8} and acoustic waves.\citep{10,11,12,13,14} WSLs and BOs were also observed in elastic systems for torsional \citep{15, 16} and surface \citep{17} waves. However, experimental studies on elastic BOs are still limited to specific experimental designs that are not easily tunable, and this fact hinders further studies and potential applications to some extent.

In this research, we numerically and experimentally demonstrate the existence of WSLs and the corresponding time-resolved BOs by using a tunable one-dimensional (1D) granular system consisting of a chain of uniform cylinders.\citep{19} According to the Hertz contact law,\citep{20} the contact stiffness between two adjacent cylinders depends on the contact angle between them, and this fact brings extreme tunability to our system. Previous researches have shown that such cylindrical granular crystals provide an enhanced freedom in tailoring stress wave propagation in solids.\citep{19,21,22a, 22b, 22c, 22d, 23,24a} For the current study, we can therefore vary the alignment of contact angles in a tunable fashion and introduce a gradient in contact stiffness along the chain---mimicking the effect of an externally applied electric field for BOs of electrons. This results in the tilting of band dispersion curves---an essential feature for the formation of the WSLs. We use a discrete element model to analyze the dynamics of the resulting system and show that it has a good agreement with the experimental observations. Furthermore, we numerically demonstrate that such systems can provide pathways for further tailoring of elastic BOs. 

\section{EXPERIMENTAL SETUP}
\begin{figure*}[h]
\centering
\includegraphics[width=6in]{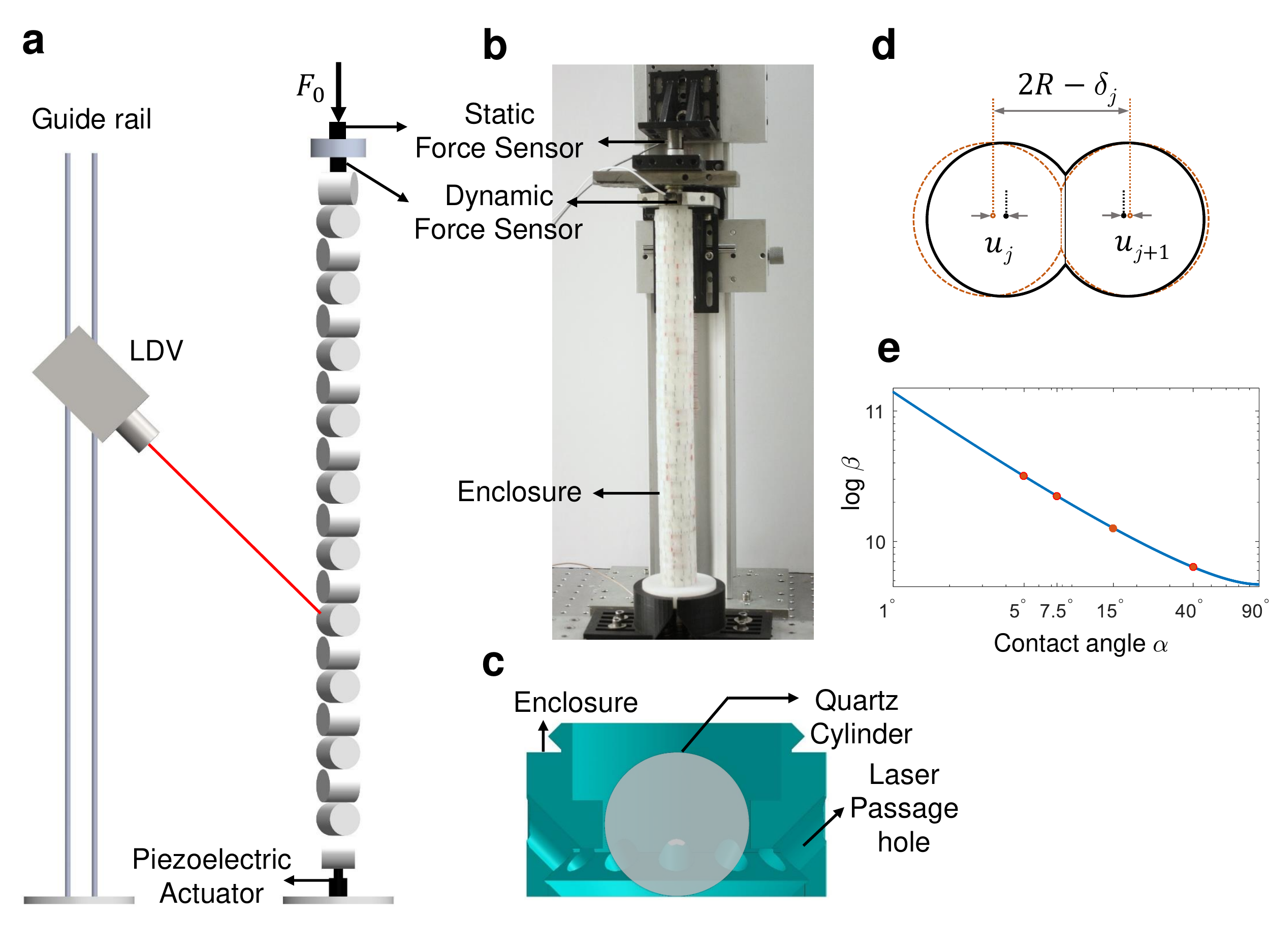}
\caption{\textbf{Experimental setup:} \textbf{(a)} The system consists of vertically stacked 1D chain of cylindrical particles under a static compression. \textbf{(b)} The actual experimental setup, in which cylinders are arranged inside 3D-printed enclosures. \textbf{(c)} Cross-sectional view of an enclosure and a cylinder inside. \textbf{(d)} Schematic of the contact interaction and related displacements. \textbf{(e)} Dependency of contact stiffness coefficient on the contact angle. The chosen angles are marked with red dots.}
\label{fig1}
\end{figure*}
The experimental setup is shown in Fig.~\ref{fig1}. The vertical 1D chain is composed of $N=21$ cylindrical granular particles made of fused quartz with density $\rho=2187$ kg/m$^3$, elastic modulus $E=72$ GPa, and Poisson's ratio $\nu=0.17$. All the cylinders are identical with the same radius $R=9$ mm and height $H=18$ mm. We use 3D-printed enclosures (white color casing in Fig.~\ref{fig1}b and enlarged view in Fig.~\ref{fig1}c) to hold the cylinder particles and stack them vertically. All the enclosures, along with the cylinders inside, can be rotated freely about their central axis. Taking advantage of the calibration markers on the outer surface of the enclosures, we are able to achieve a fine control over the rotation angle, and thus maintain contact angles with a resolution as small as 2.5$^\circ$. 

In the present study, we manipulate the contact angles between the particles to create a stiffness gradient along the length of the chain. To this end, we first take a dimer chain, i.e., with two alternating contact angles: \{$90^\circ$, $40^\circ$\}, and then modify the same judiciously to achieve a gradient. We set the first six contact angles to be \{$90^\circ$, $5^\circ$, $90^\circ$, $7.5^\circ$, $90^\circ$, $15^\circ$\}, followed by alternating \{$90^\circ$, $40^\circ$\} until the end of the chain. Note that we count the particles from the actuation side (i.e., from the bottom of the chain in Fig.~\ref{fig1}a). The primary reason behind choosing these set of angles is to achieve a large stiffness gradient at the beginning of the chain, so that WSLs and BOs are located near the actuator for an effective excitation.

We compress the system with a static force $F_0=100$ N. We mount a static force sensor at the upper bracket to measure this force. The piezoelectric actuator at the bottom excites the system with a linear chirp signal ($5$ to $30$ kHz) for band structure estimation and 0.82 ms wide (FWHM) Gaussian-modulated sinusoidal pulse (GMSP) at a target central frequency for time-history response. We also place a piezoelectric force sensor at the top end of the chain---in direct contact with the uppermost particle---to measure the transmitted dynamic force (see Fig.~\ref{fig1}a). The maximum transmitted dynamic force is about $0.55$ N, which is approximately $0.55\%$ of the initial static compression. We use a Laser Doppler Vibrometer (LDV) to measure the velocity of each particle one by one and obtain system's response. Here, the laser passage holes designed on each enclosure (Fig.~\ref{fig1}c) allow us to point the laser at 45$^\circ$ from the vertical direction and measure a component of particle velocity. We synchronize the velocity profiles measured from all particles with respect to the actuation signals to form spatio-temporal map of BOs.

\section{MODELING} 
We model the dynamics of our system by the discrete element method,\citep{25} in which we treat the chain as a system of lumped masses interconnected by nonlinear springs. This nonlinear interaction arises from the nature of the contact between cylinders, and it is governed by the Hertz's contact law.\citep{20} If $\alpha$ be the contact angle between two adjacent cylinders, the contact area would be elliptical for $0^\circ < \alpha < 90^\circ$, and circular for $\alpha = 90^\circ $. The nonlinear force-displacement relationship is thus given as\citep{21}
\begin{equation} \label{eq:er1}
F = \beta(\alpha) {z ^{3/2}},
\end{equation}
\noindent where $F$ is the contact force, and $z=\delta+\Delta{u}$ is the total relative displacement of two adjacent cylindrical particles, and it comprises of the initial static compression $\delta$ and the relative dynamic displacement $\Delta{u}$ (see Fig.~\ref{fig1}d). The contact stiffness coefficient $\beta(\alpha)$ is of the following form  
\begin{equation} \label{eq:er3}
\begin{split}
\beta(\alpha)  &= \frac{{2E}}{{3(1 - {\nu ^2})}}\sqrt {\frac{R}{{\sin \left( \alpha  \right)}}} {\left[ {\frac{2}{\pi }K(\varepsilon )} \right]^{ - \frac{3}{2}}} \\
 & \times {\left\{ {\frac{4}{{\pi {\varepsilon ^2}}}\sqrt {\left[ {{{\left( {\frac{r_1}{r_2}} \right)}^2}E(\varepsilon ) - K(\varepsilon )} \right]\left[ {K(\varepsilon ) - E(\varepsilon )} \right]} } \right\}^{\frac{1}{2}}},
\end{split}
\end{equation}
where the eccentricity $\varepsilon$ of the elliptical contact area is $\sqrt{1-\left(\frac{r_2}{r_1}\right)^2}$ with $r_1$ and $r_2$ as the semi-major and semi-minor axes, respectively. $K(\varepsilon)$ and $E(\varepsilon)$ are the complete elliptic integrals of the first and second kind, respectively, and $\frac{r_1}{r_2} \approx \left(\frac{1+\cos(\alpha)}{1-\cos(\alpha)}\right)^{-\frac{2}{3}}$.

Therefore, we write the equation of motion for the $j$th cylinder in the chain as
\begin{equation} \label{eq:er4}
{m}{\ddot u_j} = {\beta _{j-1}}\left[ {{\delta _{j - 1}} + {u_{j - 1}} - {u_j}} \right]_ + ^{\frac{3}{2}} - {\beta _{j}}\left[ {{\delta _j} + {u_j} - {u_{j + 1}}} \right]_ + ^{\frac{3}{2}} - \frac{m}{\tau}\dot{u_j}.
\end{equation}
\noindent Here $u_j$ is the dynamic displacement of the $j$th particle, and $\delta_j$ is the initial static compression between the $j$th and ($j+1$)th particles due to precompressive force. $\beta _{j}$ is the simplified notation for $\beta(\alpha_j)$ used hereafter (see its dependence on the contact angle in Fig.~\ref{fig1}e). The mass of each cylinder is $m=\rho \pi R^2H$, and $[z]_+$ denotes max(0,$z$). The final term represents the viscous force that is dependent on velocity $\dot{u_j}$ and damping coefficient $\tau$. Under a large initial static compression, we neglect the effect of gravity in this short chain. This is because the additional static compression at the bottom of the chain is about $2\%$ of the applied precompression, which leads to only $1.1\%$ variation in band edge frequencies investigated later.

\subsection{Analytical model}
\begin{figure*}[!]
\centering
\includegraphics[width=5in]{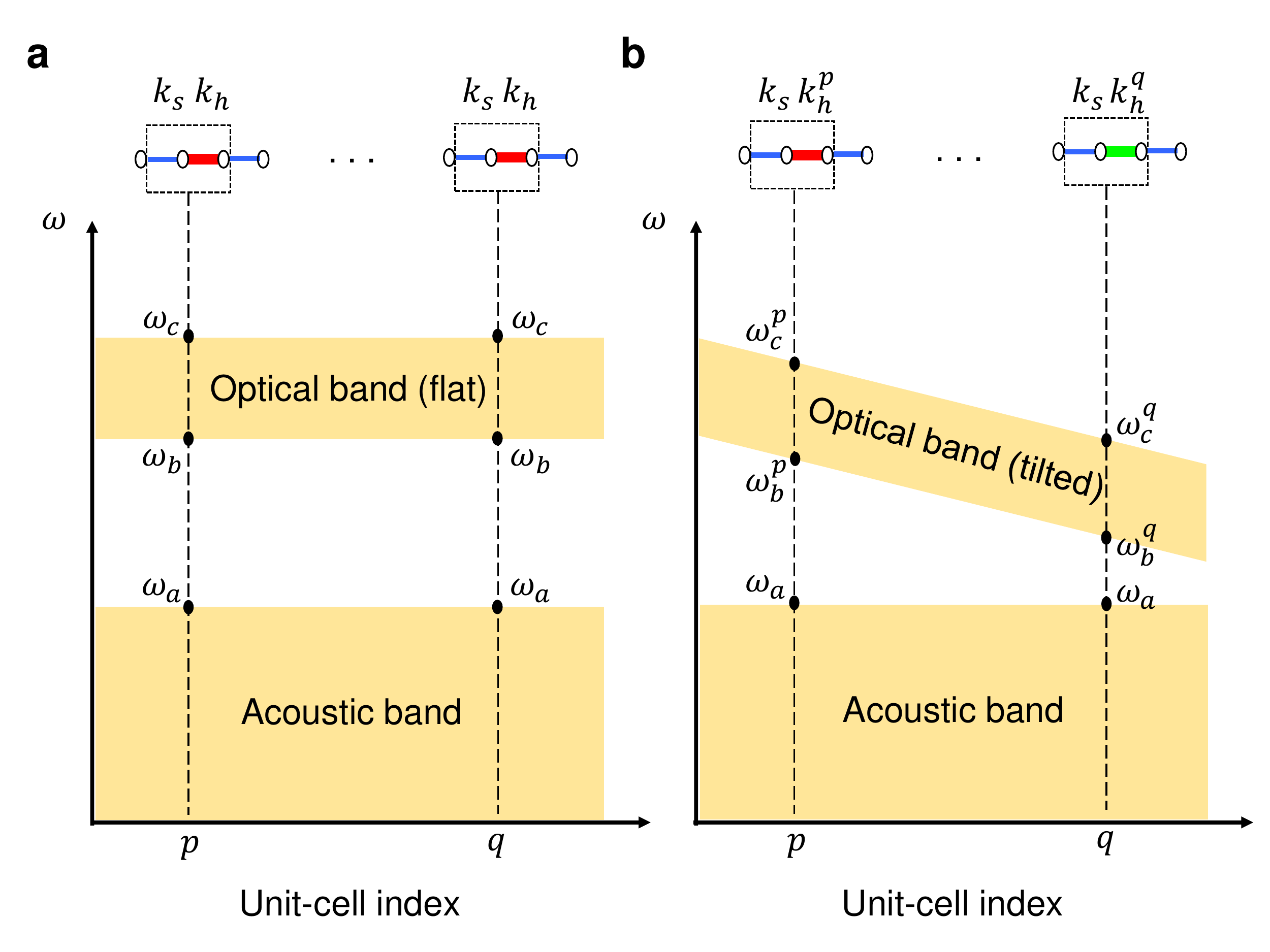}
\caption{\textbf{Spatial dependency of band dispersion curves:} \textbf{(a)} A representative dispersion of a dimer chain with flat acoustic and optical bands. \textbf{(b)} A gradient in stiffness $k_h$ leads to a tilted optical band. $p$ and $q$ are unit-cell indices for two different positions in the chain with graded stiffness values $k_h^p$ and $k_h^q$, respectively. The corresponding band edge frequencies are marked.}
\label{fig2}
\end{figure*}
To obtain the dispersion characteristics of the system, we employ an analyitcal formulation by linearizing Eq.~\eqref{eq:er4}. This is because the initial static force (displacement) is large compared to instantaneous dynamic force (displacement) in the system, therefore we have $|{u_{j+1}-{u_j}}| \ll {\delta_j}$. Moreover, neglecting the effect of viscous damping on the eigenfrequencies of our system, we suppress the damping term. We thus deduce 
\begin{equation} \label{eq:er5}
{m}{\ddot u_j} = k_{j - 1}(u_{j - 1}-u_j) - k_j(u_j-u_{j + 1}),
\end{equation}
\noindent where $k_j$ is the linearized contact stiffness between the $j$th and ($j+1$)th particles and is given as $k_j=\frac{3}{2}\beta_j\delta_j^{1/2}=\frac{3}{2}\beta_j^{1/2}F_0^{1/3}$ with $F_0$ being the initial static compression force.\citep{26} 

Now, if we have two alternating contact angles along the chain, the system is a `dimer' system with two alternating contact stiffness values, say $k_s$ (\textit{soft} contact) and $k_h$ (\textit{hard} contact). Bloch's theorem \citep{27} can be employed on this periodic system to obtain a representative dispersion curves shown in Fig.~\ref{fig2}a. This includes two bands, namely acoustic and optical bands,\citep{28} which are flat along the spatial extent of the chain, and band edge frequencies are given by\citep{28}
\begin{equation} \label{eq:er6}
{\omega _a} = \sqrt {\frac{{2{k_s}}}{m}}, \quad {\omega _b} = \sqrt {\frac{{2{k_h}}}{m}}, \quad {\omega _c} = \sqrt {\frac{{2\left( {{k_s} + {k_h}} \right)}}{m}}, \quad \quad \forall \quad k_s < k_h.
\end{equation}

Our system, however, as we have mentioned earlier, is a variant of this dimer chain, and we keep one contact angle uniform along the chain (i.e., same $k_s$) but vary the other angle to achieve a gradient in $k_h$ as a function of space---shown in the upper inset of Fig.~\ref{fig2}b. Though this system is no longer a periodic system, an approximate dispersion behavior can still be estimated by calculating the \textit{local} band edge frequencies. To this end, we take a dimer unit cell with two stiffness values at different space location and calculate band edge frequencies from Eq.~\eqref{eq:er6}. This approach is similar to the one taken in the seminal paper by Zener \cite{1}.  As a results, the optical band is tilted and becomes space dependent (Fig.~\ref{fig2}b). 

\subsection{Numerical model}
We carry out numerical simulations by directly solving the set of equations given by Eq.~\eqref{eq:er4}. To account for the boundary conditions in our finite system, we write the equation of motion for the first and last particles as
\begin{subequations}
\begin{align}
{m}{\ddot u_1} = {k _a} \left[{{\delta _{a}} + {u_0 - u_1}}\right]_ +- {\beta _1}\left[ {{\delta _1} + {u_1} - {u_{2}}} \right]_ + ^{\frac{3}{2}}-\frac{m}{\tau}\dot{u_1} \\
{m}{\ddot u_N} = {\beta _{N - 1}}\left[ {{\delta _{N - 1}} + {u_{N - 1}} - {u_N}} \right]_ + ^{\frac{3}{2}} - {k_e}\left[{{\delta _e} + {u_N}}\right]_ +-\frac{m}{\tau}\dot{u_N},
\end{align}
\end{subequations}
\noindent where $u_0$ is the displacement of the actuator, $k_a$ is the linear contact stiffness between the actuator and the first particle, and $k_e$ is the linear contact stiffness between the last particle and the dynamic force sensor, which is assumed to be fixed. The linear contact assumption is valid for high initial static compression. To match the mode shapes of the boundary modes (front and end) observed in experiments, we set the stiffness coefficients $k_a=4.2 \times 10^7$ N/m and $k_e=1.5 \times 10^8$ N/m. 
Moreover, we choose the damping factor $\tau=0.0125$ ms to match the decay of short input Gaussian pulse in the experimental results discussed later. $\delta _{a}$ ($\delta _{e}$) is the precompression between the actuator (dynamic force sensor) and the first (last) cylinder. 

We rewrite the equations in the form of first-order state space and solve them using the ODE45 solver in MATLAB. We obtain the velocity response with a sampling time step of $10^{-6}$ s. The excitation conditions are linear chirp signal and GMSP for band structure and time-history analyses, respectively, given at the front of the chain in the form of the actuator displacement $u_0$ in Eq.~(6a). 

\section{RESULTS AND DISCUSSION}
\subsection{Wannier-Stark Ladders}
\begin{figure*}[h]
\centering
\includegraphics[width=6in]{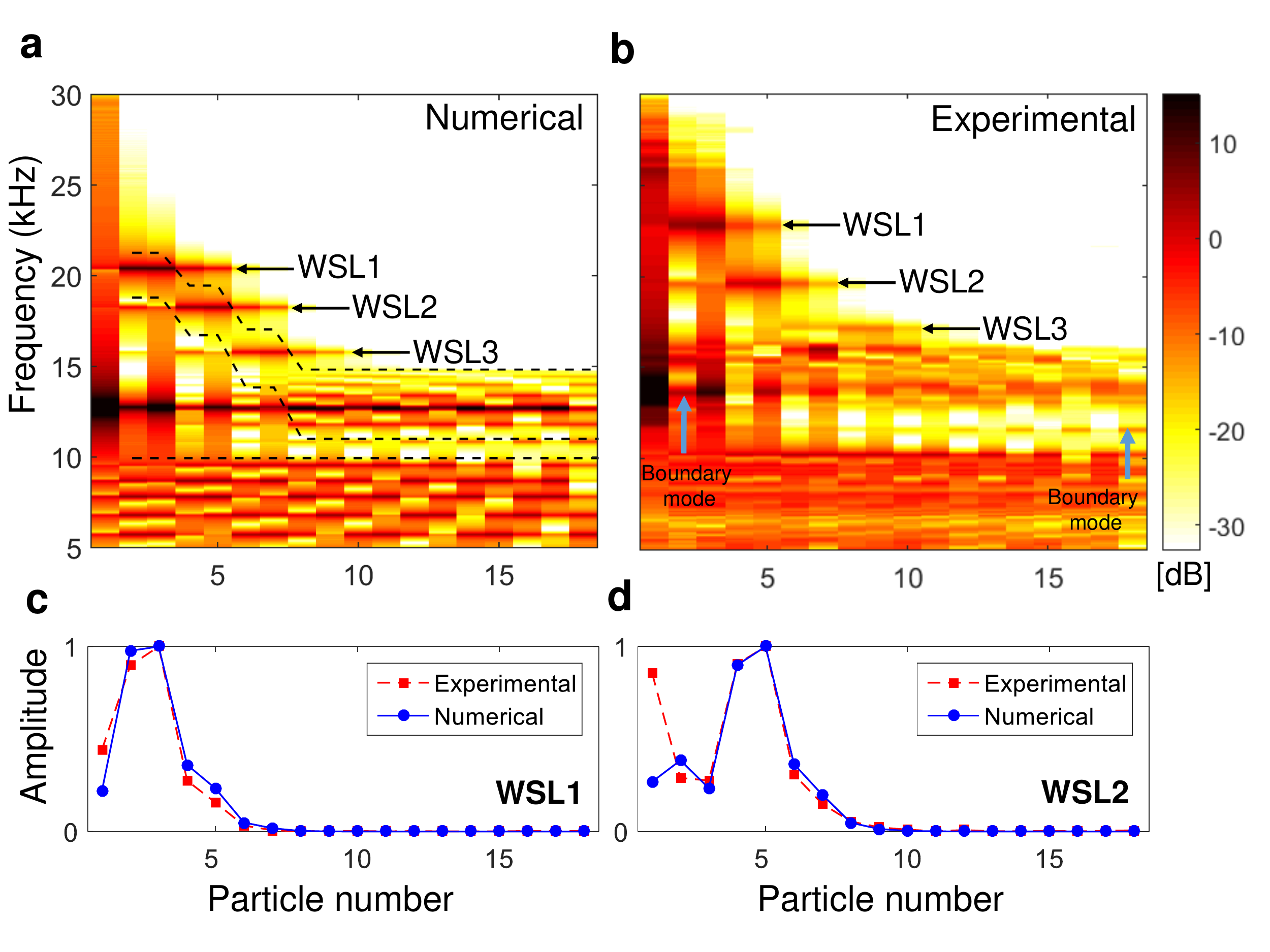}
\caption{\textbf{Spectrum along the length of the chain:} \textbf{(a)} Numerical simulations confirm the presence of titled optical band and WSLs within. Dashed black lines are analytically obtained band edge frequencies. \textbf{(b)} Experimental results corroborate the presence of the WSLs. The color bar indicates PSD obtained from velocity measurements. Boundary modes are also shown. Their profiles are used to calculate boundary stiffness values. \textbf{(c)}-\textbf{(d)} Mode shapes (absolute values) for WSL1 and WSL2 extracted from the numerical and experimental spectrum maps above.}
\label{fig3}
\end{figure*} 
WSLs are the frequency-domain counterpart of BOs. To directly observe those, we calculate the frequency band structure of the system under a chirp excitation by performing the Fast Fourier Transformation (FFT) on time-history response at each cylinder location. The power spectral density (PSD) of velocity response obtained by direct numerical simulation is shown in Fig.~\ref{fig3}a. Colored areas denote the presence of energy and the white area indicates the presence of local band gaps. For the chosen contact angles in the current setup (detailed in Section II), we observe a titled optical band in the beginning of the chain, which matches well with the analytically obtained local band edge frequencies [cf. Eq.~\eqref{eq:er6}] shown as dashed black lines.    
Within this inclined optical band, we observe WSLs in the from of three localized resonance peaks, marked as WSL1 (20.42 kHz), WSL2 (18.26 kHz), and WSL3 (15.79 kHz). 
In Fig.~\ref{fig3}b, we show experimentally obtained spectrum. We remarkably notice the presence of all three WSLs along with the titled optical band. These are WSL1 (22.80 kHz), WSL2 (19.60 kHz), and WSL3 (17.20 kHz) with relative errors of $11.66\%$, $7.34\%$, and $8.93\%$, respectively. Though we see the experimental results match numerics to a reasonable accuracy, an up-shift of WSL frequencies is observed in the experiments. This may be due to several factors, including (1) inaccuracy in the standard values of material properties, (2) deviation in the contact law exponent, and (3) stiffening caused by viscoelastic effects.\citep{g4, 22a}
In Figs.~\ref{fig4}c-d, we plot the mode shapes at the frequencies corresponding to WSL1 and WSL2, respectively. We observe that both numerical and experimental results agree well in terms of predicting the energy localization (high amplitude) at particles $\{2,3\}$ and particles $\{4,5\}$ for WSL1 and WSL2, respectively. This localization will be utilized in the next section for exciting BOs at different spatial locations depending on the input frequency.

\subsection{Time-resolved Bloch Oscillations}
\begin{figure*}[t]
\centering
\includegraphics[width=7in]{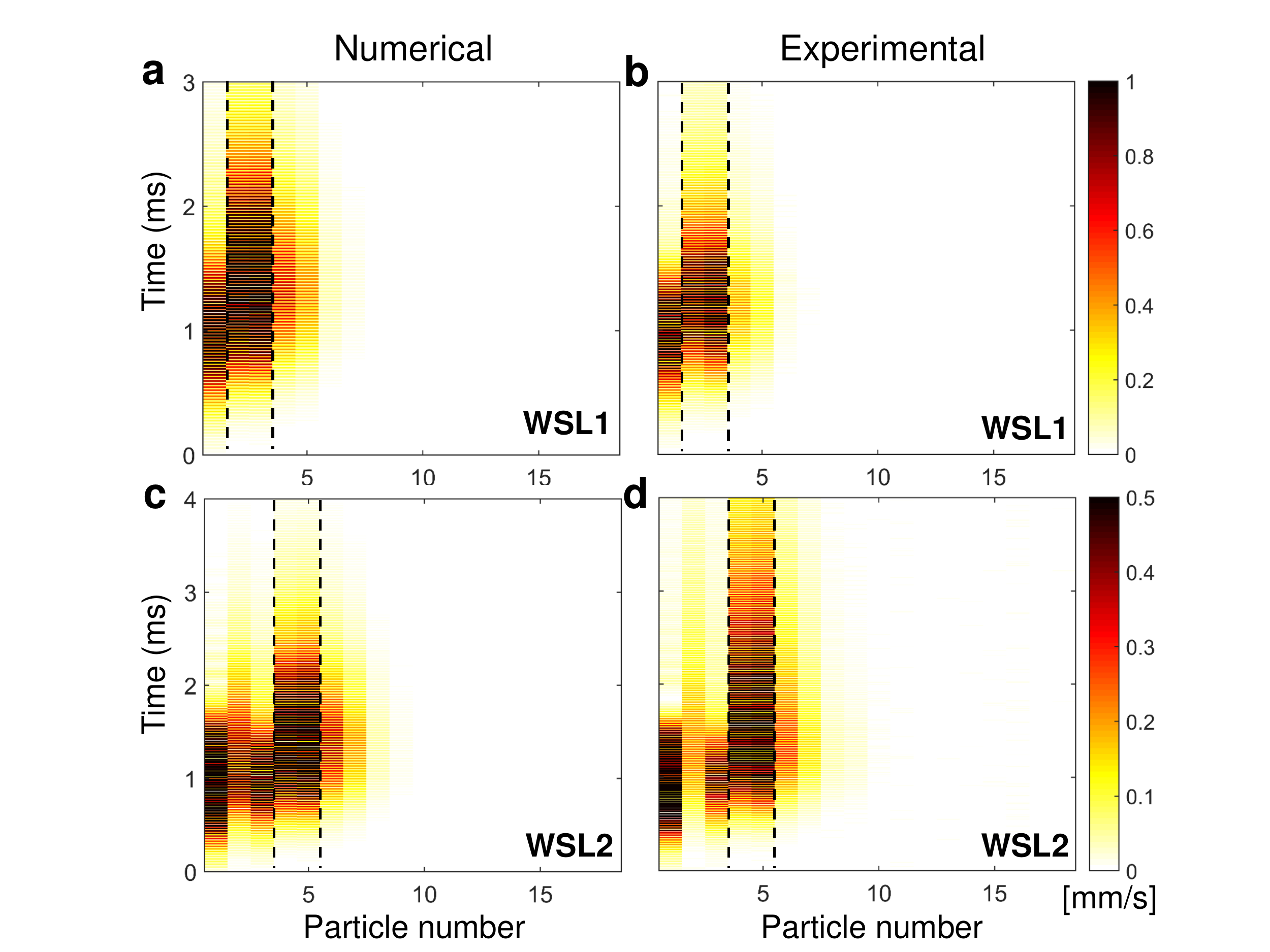}
\caption{\textbf{Time-resolved BOs and frequency-dependent energy localization:} \textbf{(a)}-\textbf{(b)} Numerical and experimental results, respectively, for the GMSP excitation centered at WSL1 frequency. \textbf{(c)}-\textbf{(d)} The same for the input excitation centered at WSL2 frequency. Dashed black lines indicate the region of localization. Color map indicates the absolute velocity of cylinders.}
\label{fig4}
\end{figure*}

To observe BOs in the time domain, we measure time-history of the velocity at each particle under the GMSP excitation centered at a WSL frequency. Figures~\ref{fig4}a-b show numerical and experimental time-resolved BOs at WSL1 frequency. We observe that as this short pulse is injected into the system, it does not travel deep and is primarily localized at the front end of the chain. Specifically, we observe more lasting vibrations at the second and third particles (enclosed within dashed black lines), and this complies well with our earlier prediction in Fig.~\ref{fig3}c that the modal energy is localized at the same spatial location in the chain. However, given the fact that we are using a short pulse with limited energy, this localized energy decays due to dissipation in the experimental setup.
We then change the central frequency of input GMSP at WSL2 and plot the results in Figs.~\ref{fig4}c-d. We clearly notice that the localization of energy is deeper into the chain---to the fourth and fifth particles. This also complies well with the localization observed in Fig.~\ref{fig3}d for WSL2. This therefore marks a direct observation of the time-resolved elastic BOs, in which we have shown a remarkable change in spatial localization of elastic energy as a function of input frequency excitation in the same system. 

\subsection{Tailorable Bloch Oscillations}
\begin{figure*}[h]
\centering
\includegraphics[width=7in]{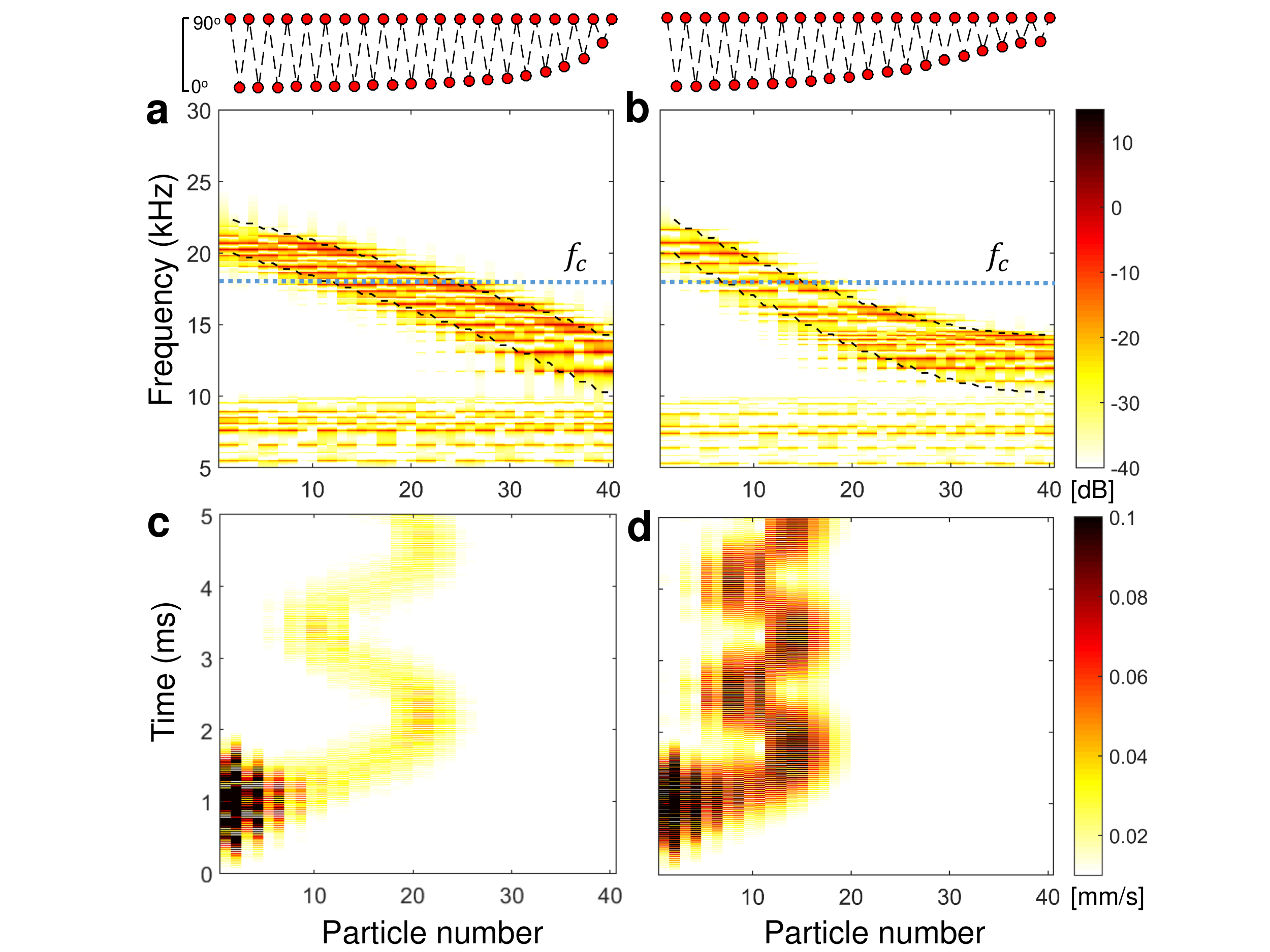}
\caption{\textbf{Tailoring BOs for the same input frequency:} \textbf{(a)} Linearly tilted optical band in spectrum plot. \textbf{(b)} Quadratically tilted optical band. The corresponding setup of cylinder contact angles is shown above. \textbf{(c)} Time-resolved localization between $10$th to $21$th particles corresponding to \textbf{(a)} for an input frequency $f_c$. \textbf{(d)} Localization occurs between $7$th to $15$th particles at the same operating frequency. Color map indicates the absolute velocity of cylinders.}
\label{fig5}
\end{figure*} 
In this section, we show how this contact-based granular chain can be further tuned to change the dispersion characteristics of the system and achieve desired spatial localization at a single operating frequency $f_c$. We numerically show that the contact angles in the chain can be fine-tuned in such a way that a linear (Fig.~\ref{fig5}a) and quadratic (Fig.~\ref{fig5}b) tilt in the optical band is obtained, which again, matches well with analytically obtained curves (dashed black lines). Here, the terms \textit{linear} and \textit{quadratic} refer to the approximate shape of the lower boundary of the optical band. The contact angle distributions along the chain to achieve such gradients of the optical band are shown in the upper insets. These are back calculated from the desired stiffness distribution by fitting a polynomial to the contact law given by Eq.~\eqref{eq:er3}. We note that this includes a minute variation of angles as small as $0.3^{\circ}$. This fine control of contact angles was not possible with the current experimental setup. Thus, we limit our investigation only to numerical study now, while we postulate that experimental verification may be possible in the future with an improved design. We send a GMSP from the front end at a center frequency $f_c=18$ kHz. Time-history plots of velocity in Figs.~\ref{fig5}c-d show distinct spatial and temporal characteristics of BOs. More specifically, for the linear optical band case, we witness the wave localization between the 10th and the 21th particles, while for the quadratic case, the localization spot has shifted to the space between the 7th and the 15th particles. Similarly, the time period of the BOs has changed from $2.43$ ms to $1.56$ ms (by comparing Figs.~5c and 5d). Lesser amplitude of velocity observed in Fig.~\ref{fig5}c is due to the fact that the localization region is farther away from the actuator, and only a limited amount of energy is pumped to excite the localized mode using evanescent waves. Therefore, we confirm that by changing the contact angles in this type of granular system, we can shift BO in space and change its period in time.

\section{CONCLUSIONS}
In this work, we have studied wave localization phenomena in a 1D granular chain consisting of vertically stacked cylindrical particles. By tuning the contact angles between the cylinders, we introduced a gradient in contact stiffness along the chain, which resulted in a spatially tilted optical band and formed the Wannier-Stark Ladders (WSLs).  We used the frequencies of these WSLs to excite spatially localized resonant modes and observed their corresponding time-resolved elastic Bloch Oscillations (BOs) through numerics and experiments. Using numerical experiments, we also showed the possibility of further tailoring these BOs in terms of altering their spatial and time responses if a more accurate experimental setup can be designed. We anticipate that this study triggers further studies on WSLs and BOs, especially in the context of nonlinear effects, which can easily be invoked in the current system. Moreover, such localized modes might be useful for engineering applications, for example, in energy harvesting using piezoelectric materials, where these resonances could potentially enhance mechanical-to-electrical energy conversion.

\section*{ACKNOWLEDGMENTS}
We thank Hiromi Yasuda and Hyunryung Kim at the University of Washington, Professor Eunho Kim in Chonbuk National University of Korea, Professor Chris Chong at Bowdoin College, and Professor Panayotis Kevrekidis at the University of Massachusetts Amherst for useful discussions and technical help. We would like to acknowledge the financial support from the NSF (CAREER-1553202) and the AFOSR (FA9550-17-1-0114).



\end{document}